\documentclass[11pt, a4paper,english,superscriptaddress,aps,nofootinbib]{revtex4}
\usepackage{amsmath,amssymb,graphicx}
\makeatletter
\usepackage{babel}
\usepackage[active]{srcltx}
\usepackage{graphicx,color}
\usepackage{changebar}
\usepackage{hyperref}
\hypersetup{
	colorlinks=true,
	urlcolor= blue,
	citecolor=blue,
	linkcolor= blue,
	bookmarks=true,
	bookmarksopen=false,
}
\usepackage[T1]{fontenc}
\usepackage{ae}
\usepackage[ansinew]{inputenc}
\usepackage{amsmath}
\usepackage{braket}
\usepackage{mathtools}
\usepackage{slashed}
\usepackage{empheq}
\usepackage{tikz}
\usepackage{multirow}
\usetikzlibrary{decorations.pathmorphing}
\usepackage[caption = false]{subfig}

\usepackage{graphicx}
\usepackage{amsfonts}
\usepackage{amsmath}
\newcommand{\be}{\begin{equation}}
\newcommand{\ee}{\end{equation}}
\newcommand{\bea}{\begin{eqnarray}}
\newcommand{\eea}{\end{eqnarray}}

\begin{document}
\title{Strong gravitational lensing in a spacetime with topological charge within the Eddington-inspired Born-Infeld gravity}

\author{C. Furtado}\email[]{furtado@fisica.ufpb.br}
\affiliation{Departamento de F\'{\i}sica, Universidade Federal da 
	Para\'{\i}ba,\\
	Caixa Postal 5008, 58051-970, Jo\~ao Pessoa, Para\'{\i}ba, Brazil}

\author{J. R. Nascimento}
\email[]{jroberto@fisica.ufpb.br}
\affiliation{Departamento de F\'{\i}sica, Universidade Federal da 
	Para\'{\i}ba,\\
	Caixa Postal 5008, 58051-970, Jo\~ao Pessoa, Para\'{\i}ba, Brazil}

\author{A. Yu. Petrov}
\email[]{petrov@fisica.ufpb.br}
\affiliation{Departamento de F\'{\i}sica, Universidade Federal da 
	Para\'{\i}ba,\\
	Caixa Postal 5008, 58051-970, Jo\~ao Pessoa, Para\'{\i}ba, Brazil}

\author{P. J. Porf\'{i}rio}\email[]{pporfirio@fisica.ufpb.br}
\affiliation{Departamento de F\'{\i}sica, Universidade Federal da 
	Para\'{\i}ba,\\
	Caixa Postal 5008, 58051-970, Jo\~ao Pessoa, Para\'{\i}ba, Brazil}

\author{A. R. Soares}
\email[]{adriano2da@gmail.com}
\affiliation{Departamento de F\'{\i}sica, Universidade Federal da 
	Para\'{\i}ba,\\
	Caixa Postal 5008, 58051-970, Jo\~ao Pessoa, Para\'{\i}ba, Brazil}

\begin{abstract}
In this work we calculate the angular deflection of light in the strong field limit in two spacetimes which were previously studied within the Eddington-inspired Born-Infeld gravity (EiBI), namely, a black hole and a wormhole, both with topological charge. We show that the presence of the parameters characterizing EiBI and the topological charge promote significant changes in the angular deflection of light with respect to that one obtained in Schwarzschild spacetime. Using the expression for angular deflection in the strong field limit, we calculate the position and magnification of the respective relativistic images.
	

\end{abstract}

\maketitle
 
 \section{INTRODUCTION}
 The Eddington-Inspired Born-Infeld gravity (EiBI) is an alternative gravity model appropriate for the high energy regime, i.e., the strong field regime \cite{Deser-Gibbons,Vollick}. This theory avoids geodesic singularities, even in a classical treatment, while asymptotically it coincides with the  General Relativity (GR) \cite{reportEiBI}. The first proposal of such modified gravity theories has been performed in a (pseudo)-Riemannian scenario \cite{Deser-Gibbons}. On the other hand, the version postulated by 
 Ba\~nados and Ferreira \cite{banados2010eddington} takes into account a metric-affine (Palatini) version of EiBI in which the gravitational sector is instead being described by the metric only, is characterized by two degrees of freedom: the metric and the connection which are supposed to be independent \textit{a priori}. As it is well known in these theories,
 the new degrees of freedom of the connection are turned on by the matter sources, otherwise, in the absence of them (vacuum case), the theory reduces to GR. This study has motivated a series of works for a variety of matter sources and many interesting solutions were found \cite{varios}. In addition, it is worth to remark the most recent results found in this theory, for example, multicenter and rotating solutions \cite{Gonzalo}, stellar structure models \cite{Gonzalo2} and scalar compact objects \cite{Gonzalo3}. The simplest solutions, to the best of our knowledge, have been found in \cite{lambaga2018, Soares2020} where the authors considered a Global Monopole (GM) as the matter source.
 
  The GM is a type of a topological defect which hypothetically has been generated within phase transitions with spontaneous symmetry breaking ($SO(3) \to U(1)$) in the early universe \cite{Kibble}.  The main theoretical line of studying such objects is devoted to their gravitational field \cite{Barriola-Vilenkin}, and these studies have been done in many alternative theories of gravity, both within metric formalism \cite{varios2} and Palatini formalism \cite{Soares2019,lambaga2018}. When an ordinary black hole absorbs a GM, the corresponding space time presents a solid angle deficit, characteristic for the topological charge. One consequence is a change in the deflection of light in comparison to the ordinary Schwarzschild black hole, it has been checked in \cite{Dadhich} by using the weak field limit \cite{Wald}.
	
Gravitational lensing is an important application of GR aimed to investigate the structure of the spacetime taking place in the cases of  the most varied gravitational sources \cite{varios3}. It can be studied either in the weak field regime when the radiation is deflected too far from the source or in the strong field regime when the radiation passes very closely to the  photon sphere and wraps several times around it. Given the mathematical  difficulties imposed by the strong field approach, the weak field scheme is the most usual  one which had achieved its success despite being mathematically simpler. However, due to earlier and more recent theoretical and experimental challenges faced by the GR, for example, one can remark the recent pictures taken from a black hole published by the Event Horizon Telescope (EHT) collaboration \cite{Akiyama:2019cqa, Akiyama:2019brx, Akiyama:2019sww, Akiyama:2019bqs, Akiyama:2019fyp, Akiyama:2019eap},  the strong field regime has acquired great importance. The reason is that  it allowed to investigate the regions localized very closely to the event horizon of a black hole or to the throat of a wormhole. Moreover, it follows from the observations that there is a possibility of detecting deviations with respect to GR, which can allow to gather information on the possible new degrees of freedom coming from the strong field corrections (high energy regime) to GR.
 
Although this problem is rather complicated, in recent years many studies have provided a reliable mathematical apparatus to work with the deflection of light in the  strong field limit \cite{Bozza2002,Tsukamoto2017}. Many applications of these studies have also been carried out. One can highlight \cite{Virbhadra-Ellis-2000}, where the authors provided, at the first time, a lens equation for a black hole setup by considering an asymptotically flat background. The remarkable feature found 
by them was the existence of an infinite number of images (besides the primary and second images) along the optic axis. The background-independent lens equation has also been found in \cite{Frittelli:1999yf}. In this scenario, the authors in \cite{Boz-Cap2001} found the angle of light deflection near the photon sphere. A new approach proposed by Tsukamoto \cite{Tsukamoto2017} allowed to improve the earlier results. The strong field limit has been explored in a variety of other contexts, for example, regular solutions (regular black holes \cite{Reg} and wormholes \cite{WH}), spinning black holes \cite{rotacao} and alternative theories of gravity \cite{TM}.

 The Barriola-Vilenkin global monopole has been investigated within the gravitational lensing context in \cite{Perlick2004}. In \cite{Man2011}, the authors obtained the impacts of the strong gravitational lensing during considering a black hole which acquired topological charge after swallowing the GM. In \cite{Sharif2015MGphantom}, the strong gravitational lensing was investigated within the contexts of an ordinary and a phantom GM. In the framework of $f(R)$ gravity, the work \cite{f(R)Man2015} also investigates strong gravitational lensing. In  \cite{Wei2015,Sotani2015} the strong field lensing by the asymptotic flat electrically charged EiBI black hole was studied. The aim of this work is to investigate the strong gravitational lensing in the spacetime of black holes and wormholes in EiBI gravity \cite{Soares2020}.

 
 The paper is organized as follows: In the section \ref{2} we obtain the metric describing the spacetime with the topological charge of the GM in EiBI gravity. In the section \ref{3} we find the deflection of light in spacetime of the black hole in the strong field limit. In the section \ref{4} we study the lensing equations and evaluate the observables related to relativistic images. In the section \ref{5} we get the expansion for deflection of light in the spacetime of the wormhole solution obtained in this theory in the strong field limit. Finally, in the section \ref{6}, we provide a summary and conclusions of the results. Throughout this paper, we use the system of units where Newton gravitational constant and the speed of light in vacuum are set equal to unity: $G=c=1$.

   \section{GLOBAL MONOPOLE SOLUTION IN E{\lowercase{i}}BI GRAVITY}\label{2}
In this section we briefly put forward the solutions of EiBI gravity with a GM taken as matter source, which have been discussed in details in \cite{lambaga2018,Soares2020}. Let us start by writing down the EiBI gravity
   \begin{equation}\label{acao-BI}
   S_{BH}=\frac{1}{\kappa^2\epsilon}\int\left[ \sqrt{-|g_{\mu\nu}+\epsilon R_{\mu\nu}(\Gamma)|}-\lambda\sqrt{-|g_{\mu\nu}|}\right]d^4x+S_m[g_{\mu\nu}, \Phi] \ ,
   \end{equation} 
 where $\epsilon$ is a constant with the mass dimension (-2). The vertical bars denote the matrix determinant. Here, the Ricci tensor $R_{\mu\nu}(\Gamma)$ is constructed on the base of the connection $\Gamma$ which, within the Palatini approach, is independent of the metric tensor $g_{\mu\nu}$. 
	The matter content is only coupled to the metric in agreement to the equivalence principle. As a result the action $S_{m}[g_{\mu\nu},\Phi]$ depends on the metric and the matter fields, denoted here as $\Phi$. The constant $\lambda$, in general, is defined as $\lambda=1+\epsilon\Lambda $, where $\Lambda$ is the cosmological constant, however, from now on, it will be neglected, i.e., $\lambda=1$.  In \cite{olmo2011}, it was shown that for a matter described by the energy-momentum tensor $T^{\mu}_{\phantom{\mu}\nu}=\text{diag}\left(-\rho,-\rho, P_{\theta}, P_{\theta}\right)$ resembling an anisotropic fluid,  the gravitational field equations are solved by the following metric:
   \begin{equation}\label{metricag}
   	ds^2=-\frac{1}{1-\epsilon\kappa^2 P_{\theta}}A(x)dt^2+\frac{1}{(1-\epsilon\kappa^2 P_{\theta})A(x)}dx^2+r^2(x)\left(d\theta^2+\sin^2\theta d\phi^2\right),
   \end{equation}
   where
   \begin{equation}\label{radial}
   	x^2=r^2(1+\epsilon\kappa^2\rho)
   \end{equation}
   and 
   \begin{equation}\label{AM}
   	A(x)=1-\frac{2M(x)}{x}, \quad\text{with}\quad\frac{dM(x)}{dx}=\frac{\kappa^2r^2\rho}{2}.
   \end{equation}
  
	We now turn our attention to the key object of this paper, that is, global monopole. The  energy-momentum tensor describing the region outside of the core of the GM \cite{Barriola-Vilenkin} is $T^{\mu}_{\phantom{\mu}\nu}=\text{diag}\left(-\frac{\eta^2}{r^2},-\frac{\eta^2}{r^2},0,0\right)$, where $\eta$ stands for the energy scale of the spontaneous symmetry breaking. With this at hand, a straightforward comparison between both aforementioned energy-momentum tensors leads us to the following identification, $\rho=\frac{\eta^2}{r^2}$ and $P_{\theta}=0$. 
  
  From (\ref{radial}), it follows that
  \begin{equation}\label{rx}
  r^2=x^2-\epsilon\alpha^2,
  \end{equation}  
  where we defined $\alpha^2\equiv\kappa^2\eta^2$.
  One can note that if $\epsilon<0$, the function $r^{2}(x)$ has a minimal value given by $r_{min}=\alpha\sqrt{|\epsilon|}$ at $x$=0, see \cite{Soares2020}. This possibility suggests the existence of wormhole-like solutions, and in fact, we will show that this solution also describes an Ellis wormhole, as the mass vanishes, plus a topological charge stemming from the GM. On the other hand, if $\epsilon>0$, the function $r^{2}(x)$ does not have a minimal non-zero value, and in this case the solution will describe a black hole with a topological charge \cite{Soares2020}. 
  
 Solving (\ref{AM}), we find $A(x)=1-\alpha^2-\frac{2M_0}{x}$, where $M_0$ is an  integration constant which can be interpreted as the mass of the object. Then, the (\ref{metricag}), expressed in terms of the $x$ coordinate, becomes
  \begin{equation}\label{metrica-x}
  	ds^2=-\left(1-\alpha^2-\frac{2M_0}{x}\right)dt^2+\left(1-\alpha^2-\frac{2M_0}{x}\right)^{-1}dx^2+(x^2-\epsilon\alpha^2)\left(d\theta^2+\sin^2\theta d\phi^2\right).
  \end{equation}
 We can rewrite this line element in terms of the $r$ coordinate by means of (\ref{rx}). For the region  corresponding to $x>0$ and requiring $\epsilon>0$, the metric is
  \begin{equation}\label{metrica-r}
  ds^2=-\left(1-\alpha^2-\frac{2M_0}{\sqrt{r^2+\epsilon\alpha^2}}\right)dt^2+\frac{r^2}{r^2+\epsilon\alpha^2}\left(1-\alpha^2-\frac{2M_0}{\sqrt{r^2+\epsilon\alpha^2}}\right)^{-1}dr^2+r^2(d\theta^2+\sin^2\theta d\phi^2).
  \end{equation}
   This solution was first obtained in \cite{lambaga2018}. From  (\ref{metrica-x}), we find that picking the integration constant to be $M_0=0$, the $\epsilon<0$ case provides a solution describing an Ellis-like wormhole with topological GM charge \cite{Soares2020}. Hence, doing this and rescaling the metric (\ref{metrica-x}): $t\to t\sqrt{1-\alpha^2}$, $x\to\frac{x}{\sqrt{1-\alpha^2}}$ and $\epsilon\to\frac{\epsilon}{1-\alpha^2}$, we have
   \begin{equation}\label{wormhole}
   		ds^2=-dt^2+dx^2+(1-\alpha^2)(x^2+|\epsilon|\alpha^2)\left(d\theta^2+\sin^2\theta d\phi^2\right).
   \end{equation}
 It is worth calling attention to the fact that this solution arises naturally in the EiBI case and does not violate the energy conditions. In addition, it is the simplest solution of a wormhole in EiBI gravity that we have known up to now. 
   
   For the sake of simplicity, we will  suppose all distances to be measured in terms of the Schwarzschild mass: $x\to \frac{x}{2M_0}, t\to \frac{t}{2M_0}, \epsilon\to \frac{\epsilon}{(2M_0)^2}$. Thus the metric (\ref{metrica-x}) takes the form
   \begin{equation}\label{mtr}
   		ds^2=-(1-\alpha^2-\frac{1}{x})dt^2+\frac{dx^2}{(1-\alpha^2-\frac{1}{x})}+(x^2-\epsilon\alpha^2)(d\theta^2+\sin^2\theta d\phi^2) \ .
   \end{equation}
The methodology for treating gravitational lensing we are going to adopt in the following cannot be applied to naked singularities. So, let us restrict ourselves to the branch $x>0$ in the former equation. We should also point out that in the case $\epsilon<0$, we can have $0<x<\infty$ and there is no possibility for arising naked singularities. Already in the case $\epsilon>0$, we have $\alpha\sqrt{|\epsilon|}<x<\infty$; therefore, from a direct inspection of the former equation one concludes that the horizon is located at $x_{h}=\frac{1}{1-\alpha^2}$ and, as a result of (\ref{rx}),  $x_h$ must be bigger than the lowest value of the  coordinate $x$, that is, we must have \cite{lambaga2018}
\begin{equation}\label{rest}
\alpha^2\epsilon\le\frac{1}{(1-\alpha^2)^2},
\end{equation}
to ensure $r$ real.

In the next section we shall investigate the  deflection of light very closely these objects. For that, we shall implement the strong field scheme which is the most suitable approach to dealing with black holes and wormholes as remarked before.

\section{STRONG FIELD DEFLECTION OF LIGHT}\label{3}
 In order to introduce the main elements for the calculation of light deflection in the strong field limit, let us follow the methodology introduced by Bozza \cite{Bozza2002} and improved by Tsukamoto \cite{Tsukamoto2017}. We start by defining a generic static and spherically symmetric spacetime whose line element is
\begin{equation}\label{gm}
	ds^2=-A(x)dt^2+B(x)dx^2+C(x)\left(d\theta^2 +\sin^2\theta d\phi^2\right) \ .
\end{equation}
A photon starting from infinity \footnote{ In the case of spacetime with a topological charge, we will have the limits $\lim\limits_{x\to\infty}A(x)=1-\alpha^2$, $\lim\limits_{x\to\infty}B(x)=(1-\alpha^2)^{-1}$ and $\lim\limits_{x\to\infty}C(x)=x^2-\epsilon\alpha^2$. With this we can show that the effective potential $V$ satisfies the condition $\lim\limits_{x\to\infty}V=\left(\frac{E}{1-\alpha^2}\right)^{2}$. As the motion of the photon is allowed in the region $V\ge0$, the topological charge does not prevent the asymptotic existence of the photon.}, when it approaches a black hole with a given parameter of impact $b$, is deflected at a closest approach $x_0$ and then goes to infinity. The impact parameter is related to the closest approach through the relation 
\begin{equation}
	b=\sqrt{\frac{C_0}{A_0}} \ ,
\end{equation}
where the subscript ``0" indicates that the function  is evaluated at $x=x_0$, i.e., $A_0=A(x_0)$. 
It can be shown that the deflection of light, $\Lambda(x_0)$, is expressed in terms of the closest approach $x_0$ as
\begin{equation}\label{d}
	\Lambda(x_0)=I(x_0)-\pi,
\end{equation}
where
\begin{equation}
	I(x_0)=2\int_{x_0}^{\infty}\frac{dx}{\sqrt{\frac{R(x)C(x)}{B(x)}}}\quad\text{and} \quad R(x)=\frac{A_0C}{AC_0}-1.
\end{equation}
At the weak field limit this integral is expanded up to the first order in $\epsilon$ and the mass $M$ \cite{lambaga2018, Soares2019}. When $x_0$ is strongly different from $b$, this limit is no longer valid. In this case, we must consider the strong field limit. The dependence behaves as follows: the lower $x_0$, the greater the deflection. Then when $x_0$ coincides with the radius of the photon sphere $x_{m}$  the deflection angle diverges. In \cite{Bozza2002} it was demonstrated that this divergence is logarithmic, in addition, there was presented an algorithm for the calculation of the angular deflection. Recently, Tsukamoto \cite{Tsukamoto2017} presented a further development of the method presented in \cite{Bozza2002},  and this approach will be used within our calculations.

After introducing the variable 
\begin{equation}\label{v}
	z=1-\frac{x_0}{x} \ ,
\end{equation}
 $I(x_0)$ can be presented as
\begin{equation}
	I(x_0)=\int_{0}^{1}f(z,x_0)dz \ ;
\end{equation}
where 
\begin{equation}\label{f-G}
f(z,x_0)=\frac{2x_0}{\sqrt{G(z,x_0)}}	\quad\text{with}\quad G(z, x_0)=\frac{RC}{B}(1-z)^4 \ .
\end{equation}
The integral $I(x_0)$ can be split into two parts: the divergent one, $I_D(x_0)$ and the regular one, $I_R(x_0)$. The divergent part is defined by
\begin{equation}
	I_D(x_0)=\int_{0}^{1}f_D(z,x_0)dz.
\end{equation}
where $f_D(z,x_0)=\frac{2x_0}{\sqrt{c_1z+c_2z^2}}$,  $c_1$ and $c_2$ stand for coefficients of the power series expansion of the function $G(z,x_0)$ up to the second order in $z$ (\ref{f-G}). To find the regular part one only needs to subtract the divergent part from $I(x_0)$ so that
\begin{equation}\label{regp}
	I_R(x_0)=\int_{0}^{1}f_R(z,x_0)dz, \quad f_R(z,x_0)=f(z,x_0)-f_D(z,x_0).
\end{equation}

In the strong field limit ($x_0\to x_m$, or $b\to b_c$), the deflection angle of the light (\ref{d}) is given by
\begin{equation}\label{angdefle}
	\Lambda(b)=-\bar{a}\log\left(\frac{b}{b_c}-1\right)+\bar{b}+\mathcal{O}[(b-b_c)\log(b-b_c)] \ ;
\end{equation}
 where 
\begin{equation}\label{bcc}
	 b_c=\lim\limits_{x_0\to x_m}\sqrt{\frac{C_0}{A_0}} \ ,
\end{equation}
\begin{equation}
\bar{a}=\sqrt{\frac{2B_mA_m}{C''_mA_m-C_mA''_m}} \ ,
\end{equation}
and 
\begin{equation}\label{bbarra}
	\bar{b}=\bar{a}\log\left[x_m^2\left(\frac{C''_m}{C_m}-\frac{A''_m}{A_m}\right)\right]+I_R(x_m)-\pi \ .
\end{equation}
  The superscript ${\bf X}''_m$ means second derivative of ${\bf X}(x)$ with respect to  $x$ evaluated in $x=x_m$, i.e., ${\bf X}''_m=\frac{d^2\mathbf{X}(x)}{dx^2}\Big|_{x=x_m}$. 
The next step is to calculate the coefficients of expansion, the radius of the photon sphere $x_m$, the critical parameter $b_c$,  and also, $\bar{a}$ and $\bar{b}$. We note that although the geometry we are going to study differs slightly from Schwarzschild one, the mathematical expressions describing the quantities mentioned above will become much more complicated than Schwarzschild ones. Thus, we will present the analytical expressions in first order in the EiBI parameter while keeping the part relative to the GR in an exact form. We will also present the results numerically from the quantities without approximation. We show, in comparison with the relativistic case, what changes occur due to the parameter $\epsilon$.

\subsection{Light deflection in EiBI GM}
\begin{figure}[h]
 	\centering
 	\includegraphics[height=5cm]{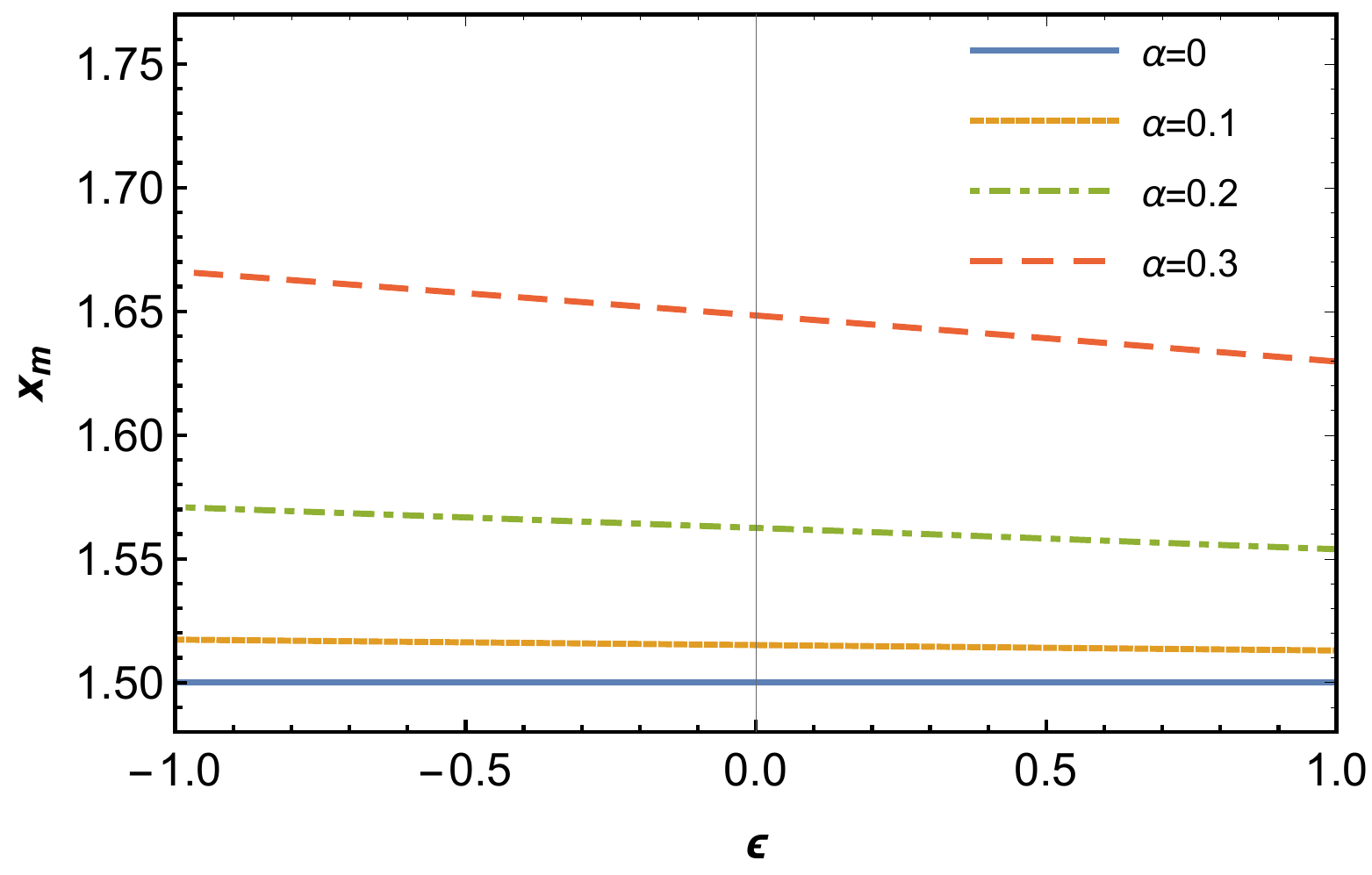}
 	\caption{ Radius of the photon sphere for several values of $\alpha$, in function of $\epsilon$.} 
 	\label{xm}
 \end{figure}

 Our next step is obtaining the analytical expression for the deflection of light in EiBI GM metric  (\ref{mtr}) in the strong field limit.
 Comparing (\ref{mtr}) and (\ref{gm}), we have
 \begin{equation}\label{AC}
 	A(x)=\frac{1}{B(x)}=1-\alpha^2-\frac{1}{x}\quad\text{and}\quad C(x)=x^2-\epsilon\alpha^2 \ .
 \end{equation}
As found in \cite{Bozza2002}, the radius of the photon sphere for a static spherically symmetric spacetime is the largest of the real solutions of the following equation:
 \begin{equation}\label{rps}
 	\frac{C'(x)}{C(x)}-\frac{A'(x)}{A(x)}=0 \ .
 \end{equation}
Substituting (\ref{AC}) into (\ref{rps}), we arrive at
 \begin{equation}\label{ce}
 	x^3-\frac{3x^2}{2(1-\alpha^2)}+\frac{\epsilon\alpha^2}{2(1-\alpha^2)}=0 \ .
 \end{equation}
 If we take $\epsilon=0$, we readily get the real root $x_m=\frac{3}{2(1-\alpha^2)}$, which is the result predicted by GR \cite{Man2011}. To find $x_m$ in the EiBI gravity  we must solve the cubic equation (\ref{ce}). The exact solution of (\ref{ce}) is given in the Appendix \ref{ap}. In the leading order, we have
 \begin{equation}\label{xaproximado}
 		x_{m}\simeq\frac{3}{2(1-\alpha^2)}-\frac{2\epsilon\alpha^2}{9} \ .
 \end{equation}
 Thus, the photon sphere radius decreases when $\epsilon$ increases. At the Fig. \ref{xm}  we display the behavior of $x_m$ as a function of $\epsilon$ for some values of $\alpha$.
From Fig. \ref{xm} it becomes clearer that the radius of the photon sphere grows as $\alpha$  increases, but decreases as $\epsilon$  increases. Note that in the case of $\epsilon<0$, this means that the radius of the photon sphere increases as the $|\epsilon|$  increases. The solid line corresponds to the case of Schwarzschild solution, where $x_m=1.5$. It must be noted that this result corresponds to the interval $x>0$, and not the entire range of $x$.
 
\begin{figure}
  	\subfloat[Critical parameter $b_c$]{\includegraphics[width = 3in]{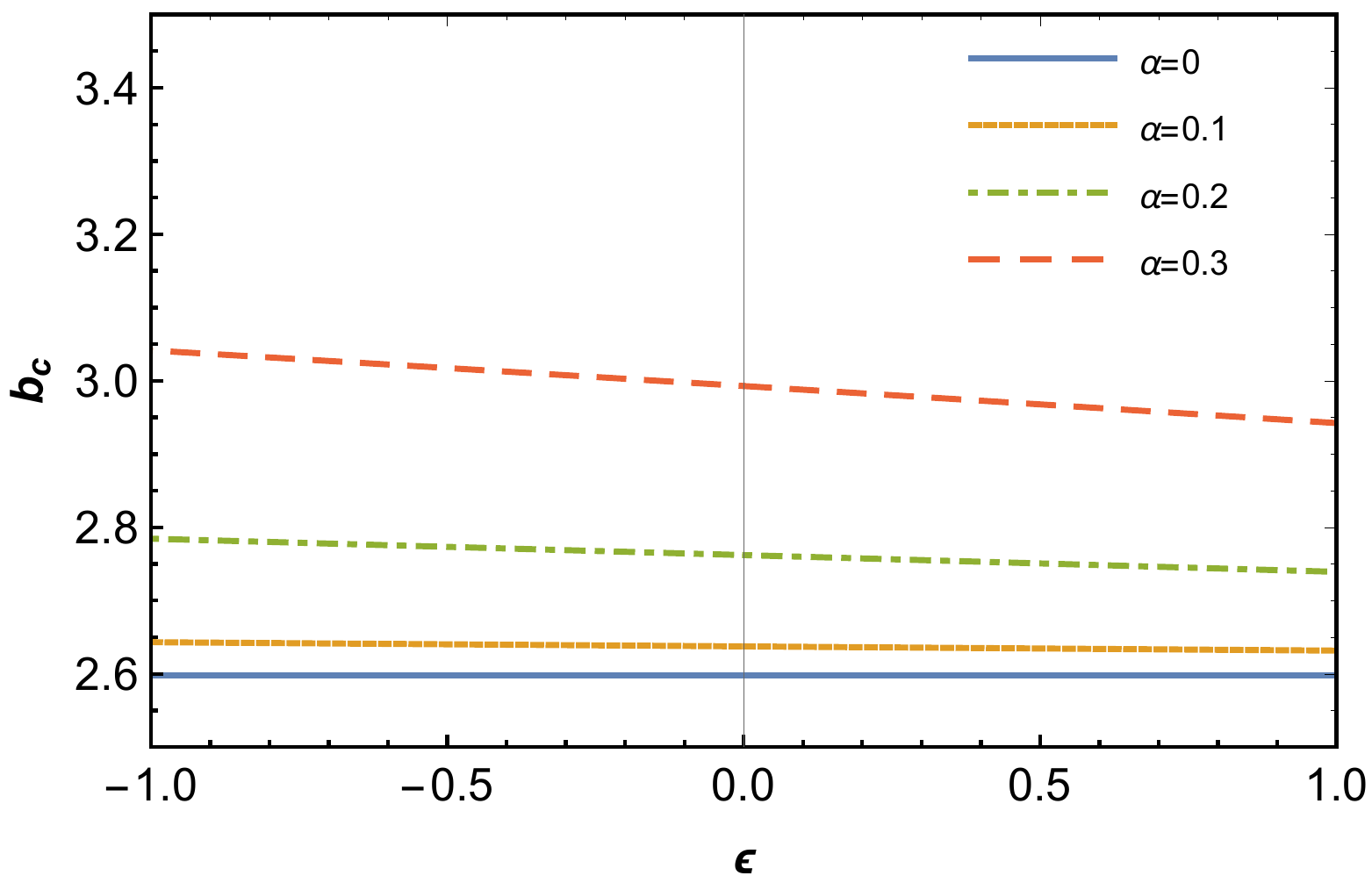}} 
  	\subfloat[$\bar{a}$]{\includegraphics[width = 3in]{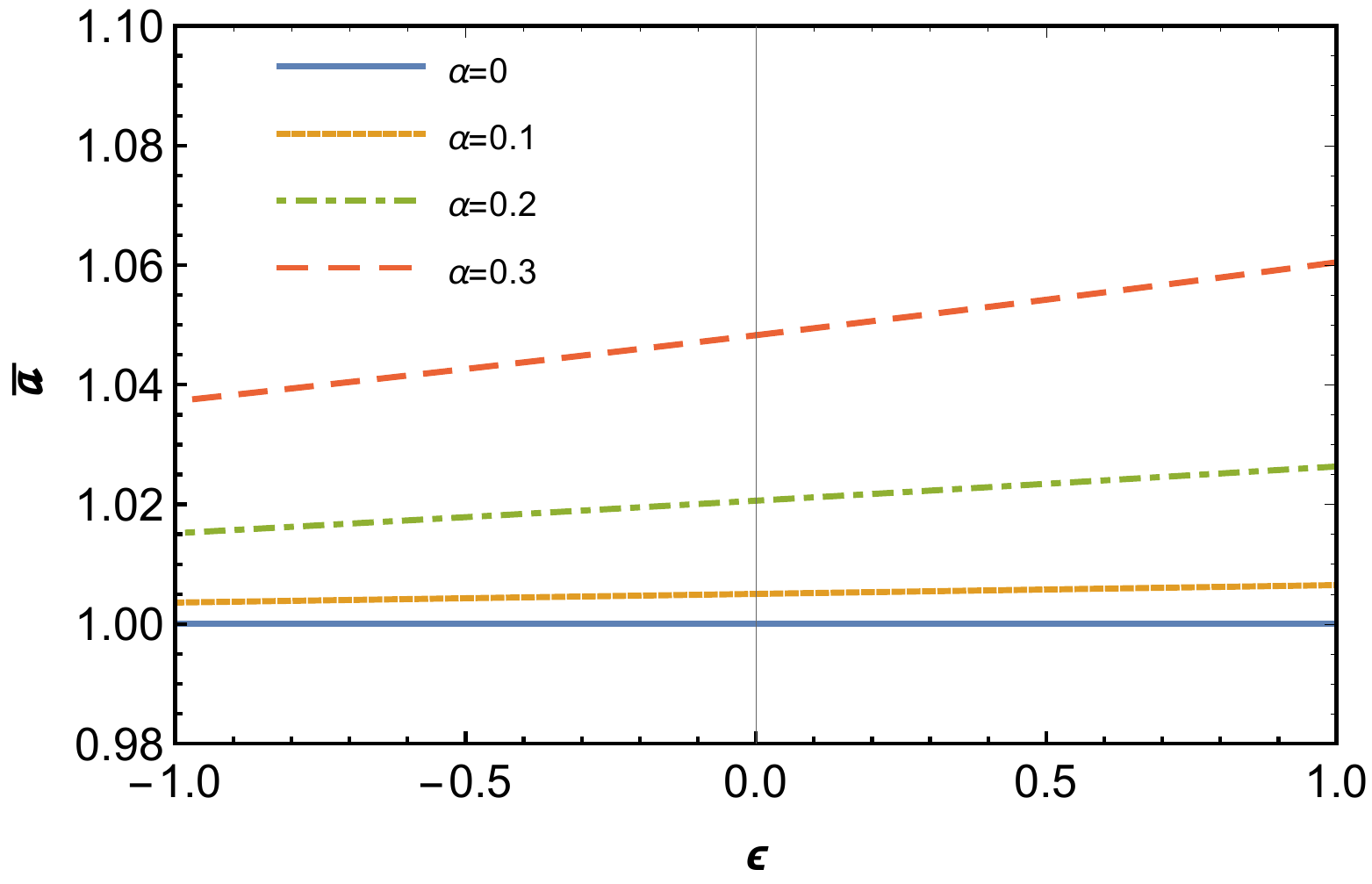}}\\
  	\subfloat[$\bar{b}$]{\includegraphics[width = 3in]{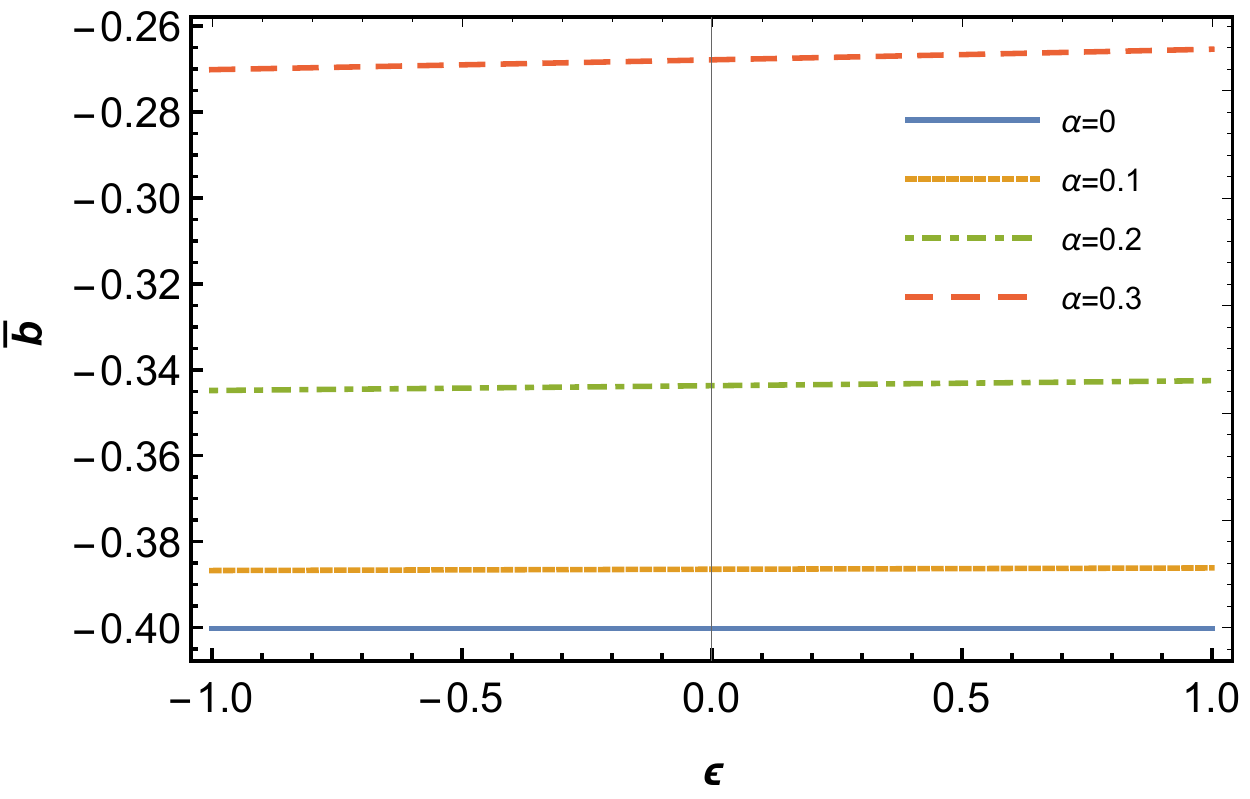}}
  	\subfloat[Deflection angle $\Lambda(\theta)$]{\includegraphics[width = 3in]{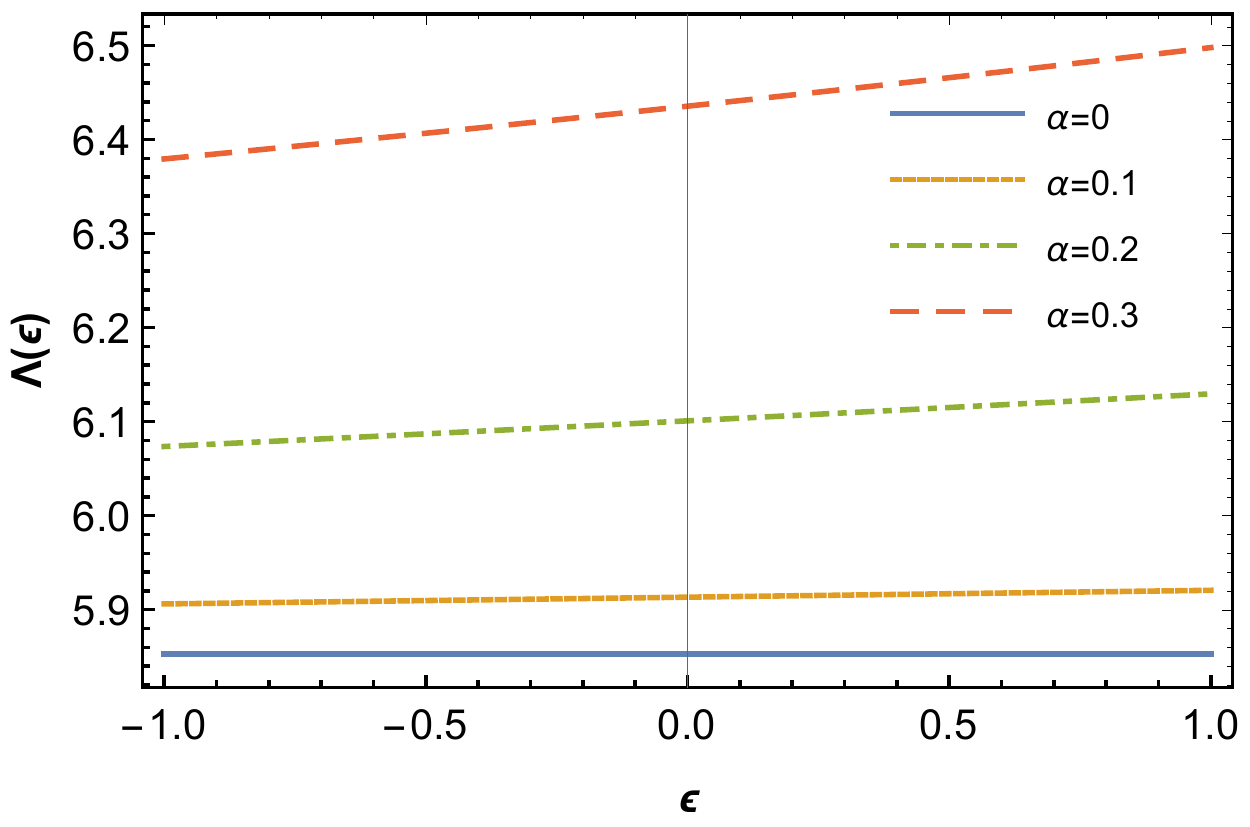}} 
  	\caption{Coefficients of the strong field gravitational lensing expansion.}
  	\label{CE}
  \end{figure}

 From (\ref{AC}) and (\ref{bcc}), the critical impact parameter $b_c$ is given by
 \begin{equation}
 	b_c=\sqrt{\frac{x_m^3-\epsilon\alpha^2x_m}{(1-\alpha^2)x_m-1}} \ .	
 \end{equation}
 In the leading order, the (\ref{xaproximado}) leads to 
 \begin{equation}\label{bcri}
 	b_c\simeq\frac{3\sqrt{3}}{2(1-\alpha^2)^{3/2}}-\frac{\epsilon\alpha^2}{\sqrt{3}} \ ,
 \end{equation}
where the first term corresponds to the GR. As we can see, the critical parameter displays the same behavior as the radius of the photon sphere, that is, it decreases when  $\epsilon$ increases, but grows when $\alpha$ increases. In the simple Schwarzschild case, $\alpha=0$, $b_c=\frac{3\sqrt{3}}{2}\approx 2.6$. We present the behavior of $b_c$ at the Fig. \ref{CE}a.
 
  Now let us look at the term $\bar{a}$, which is given by
 \begin{equation}
 	\bar{a}=\sqrt{\frac{1}{1-\alpha^2-\frac{\epsilon\alpha^2}{x_m^3}}} \ .
 \end{equation}
 In the leading order,
 \begin{equation}\label{abarra}
 	\bar{a}\simeq \frac{1}{\sqrt{1-\alpha^2}}+\frac{4\epsilon\alpha^2}{27} \ ,
 \end{equation}
 here the first term corresponds to the exact expression in the GR and the second is the first order contribution in the EiBI parameter $\epsilon$. We note  that $\bar{a}$ grows when $\epsilon$ increases and also when $\alpha$ increases. The behavior of $\bar{a}$ for several values of $\alpha$ is presented at  the Fig. \ref{CE}b.
 
  Now we obtain $\bar{b}$, defined by (\ref{bbarra}): 
 \begin{equation}\label{barra}
 		\bar{b}=\bar{a}\log\left[x_m^2\left(\frac{C''_m}{C_m}-\frac{A''_m}{A_m}\right)\right]+I_R(x_m)-\pi,
 \end{equation}
 where $I_{R}(x_m)$ is given by (\ref{regp}). Let us evaluate $\bar{b}$ step by step. First, it is convenient to define $\mathcal{I}=x_m^2\left(\frac{C''_m}{C_m}-\frac{A''_m}{A_m}\right)$, which with use of (\ref{AC}), can be rewritten as
 \begin{equation}\label{I}
 	\mathcal{I}=\frac{2x_m^2}{x_m^2-\epsilon\alpha^2}-\frac{2}{(1-\alpha^2)x_m-1} \ .
 \end{equation}
 
  Substituting (\ref{xaproximado}) in (\ref{I}) we have in the leading order: 
 \begin{equation}\label{Ia}
 	\mathcal{I}\simeq 6+\frac{8\epsilon\alpha^2}{3} \ ,
 \end{equation}
  where the first term corresponds to the GR and the second one is the EiBI leading-order correction. From (\ref{regp}), the regular integral is found to be 
    \begin{eqnarray}\label{inR}
    I_R(x_m)&=&\int_{0}^{1}\frac{2\sqrt{x_m(4x_m^3-2\epsilon)} \ dz}{\sqrt{(4x_m\epsilon-6\epsilon)z^5-(14x_m\epsilon-21\epsilon)z^4-(4x_m^3-16x_m\epsilon+22\epsilon )z^3+(6x_m^3+6\epsilon-6x_m\epsilon)z^2}}\nonumber\\
    &-&\int_{0}^{1}\frac{2\sqrt{x_m(2x_m^3-\epsilon)} \ dz}{z\sqrt{3(x_m^3-x_m\epsilon+\epsilon)}} \ ,
    \end{eqnarray}
    or, in a more compact form,
    \begin{eqnarray}
   I_R(x_m)&=&\int_{0}^{1} \frac{2x_m \sqrt{4 x_m^3-2 \epsilon } \ dz}{\sqrt{x_m z^2 (2 z-3) \left(\epsilon  \left(2x_m (z-1)^2-3 (z-2) z-2\right)-2 xm^3\right)}}\nonumber\\
   &-&\int_{0}^{1}\frac{2 x_m \sqrt{2 x_m^3-\epsilon } \ dz}{ \sqrt{3x_m z^2 \left(x_m^3-x_m \epsilon +\epsilon \right)}} \ ;
    \end{eqnarray}
    where the second term corresponds to the divergent part. Substituting (\ref{xaproximado}) into the (\ref{inR}), we are left with
    \begin{equation}\label{Iapx}
    	I_{R}(x_m)\simeq \frac{4}{\sqrt{1-\alpha^2}}\log(3-\sqrt{3})-\frac{4\epsilon\alpha^2}{27}\left(9-2\sqrt{3}-4\log\left(3-\sqrt{3}\right)\right).
    \end{equation}
    From (\ref{abarra}), (\ref{Ia}) and (\ref{Iapx}), the approximate expression for $\bar{b}$,  given by the equation (\ref{barra}), can be written as
    \begin{eqnarray}\label{bbarra1}
    	\bar{b}&\simeq& \left(\frac{1}{\sqrt{1-\alpha^2}}+\frac{4\epsilon\alpha^2}{27}\right)\log(6+\frac{8\epsilon\alpha^2}{3})\nonumber\\
    	&+&\frac{4}{\sqrt{1-\alpha^2}}\log(3-\sqrt{3})-\frac{4\epsilon\alpha^2}{27}\left(9-2\sqrt{3}-4\log(3-\sqrt{3}\right)-\pi.
    \end{eqnarray}
      The expression (\ref{barra}) can be evaluated numerically. The result is presented in Fig. \ref{CE}c. One can observe that $\bar{b}$ grows when $\epsilon$ and $\alpha$ increase. In the GR case, when $\alpha=0$, $\bar{b}\approx-0.4002$.  Finally, we can write the approximate deflection at the strong field limit of the spacetime (\ref{mtr}). Substituting (\ref{bcri}), (\ref{abarra}) and (\ref{bbarra1}) in (\ref{angdefle}) we
		 have in the leading order,
    \begin{eqnarray}
    	\Lambda(b)&\simeq&-\left(\frac{1}{\sqrt{1-\alpha^2}}+\frac{4\epsilon\alpha^2}{27}\right)\log\left(\frac{b}{\frac{3\sqrt{3}}{2(1-\alpha^2)^{3/2}}-\frac{\epsilon\alpha^2}{\sqrt{3}}}-1\right)+\left(\frac{1}{\sqrt{1-\alpha^2}}+\frac{4\epsilon\alpha^2}{27}\right)\log(6+\frac{8\epsilon\alpha^2}{3})\nonumber\\
    	&+&\frac{4}{\sqrt{1-\alpha^2}}\log(3-\sqrt{3})-\frac{4\epsilon\alpha^2}{27}\left(9-2\sqrt{3}-4\log(3-\sqrt{3})\right)-\pi.
    \end{eqnarray}
   As expected, taking $\epsilon=0$, we recover the expression for the angular deflection obtained in GR \cite{Man2011}. In  Fig. \ref{CE}d we plotted the deflection angle evaluated at $b=b_c+0.005$. One can see that the angular deflection, besides increasing with $\alpha$, as we already know, also increases with $\epsilon$. In the Schwarzschild case, $\Lambda\approx 5.8528$. Therefore, the analysis of light deflection in the strong field gravitational lensing allows us to distinguish between the black hole with topological charge predicted by GR and EiBI gravity.
   
 \section{LENS EQUATION}\label{4}
 
\begin{figure}[h]
 	\centering
 	\includegraphics[height=7cm]{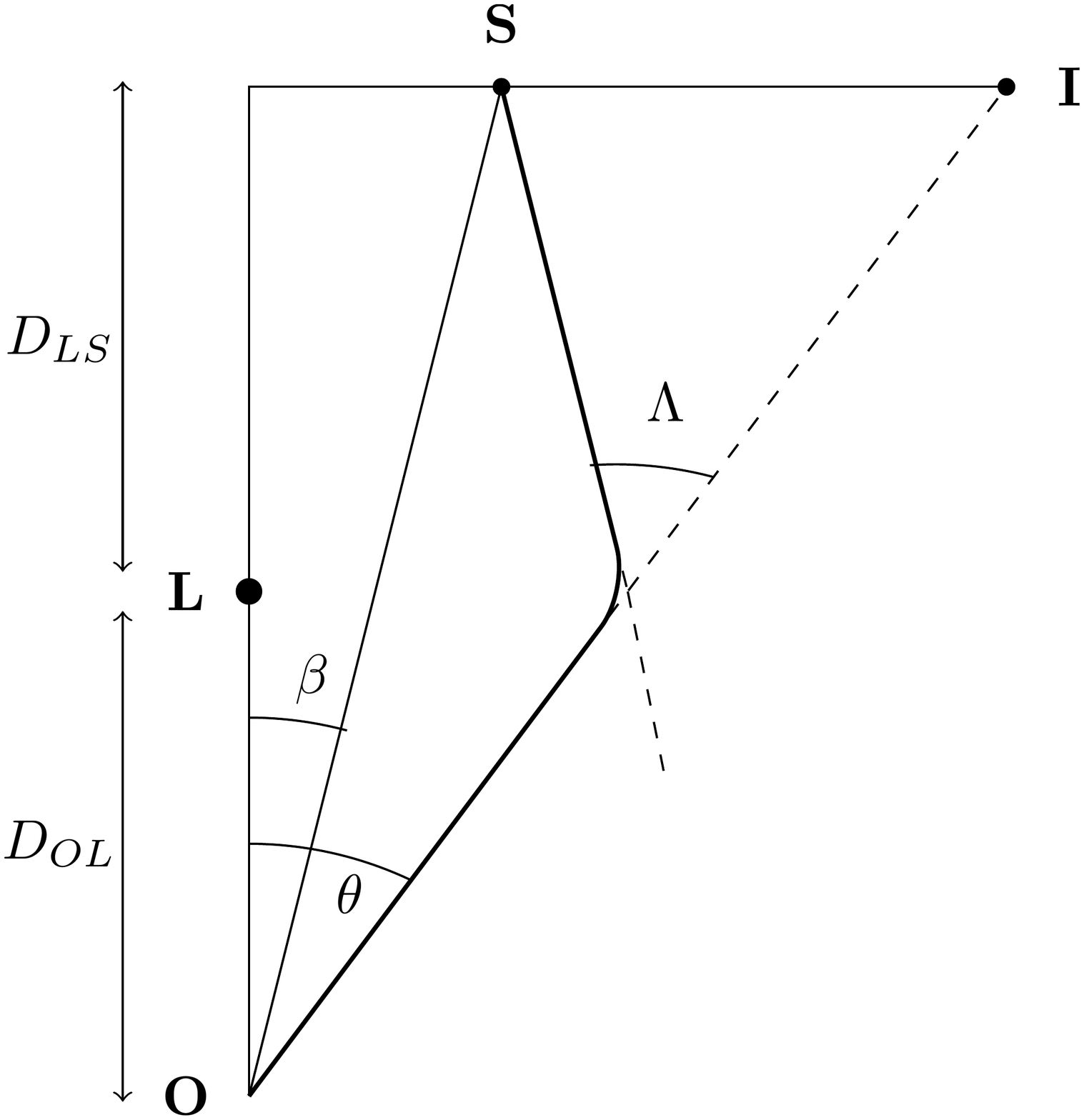}
	\caption{Lens diagram}
	\label{LG} 	
 \end{figure}

 After obtaining the angular deflection of light, we are now able to study gravitational lenses. Keeping this purpose in mind, we start this section by briefly reviewing the lens equation in the strong field limit. For that, let us consider the lens configuration outlined in Fig. \ref{LG}.


\begin{figure}
	\subfloat[$\theta_{\infty}$]{\includegraphics[width = 3in]{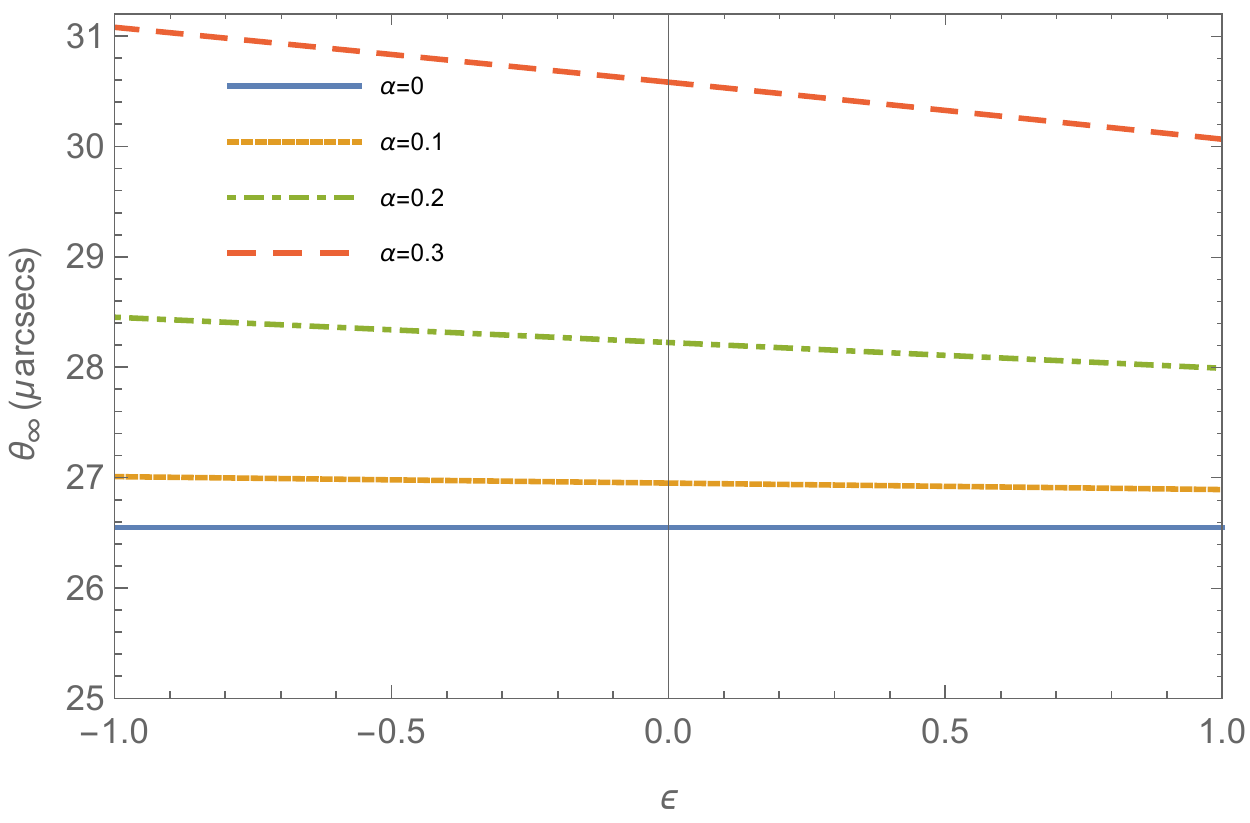}} 
	\subfloat[Angular separation $s$]{\includegraphics[width = 3in]{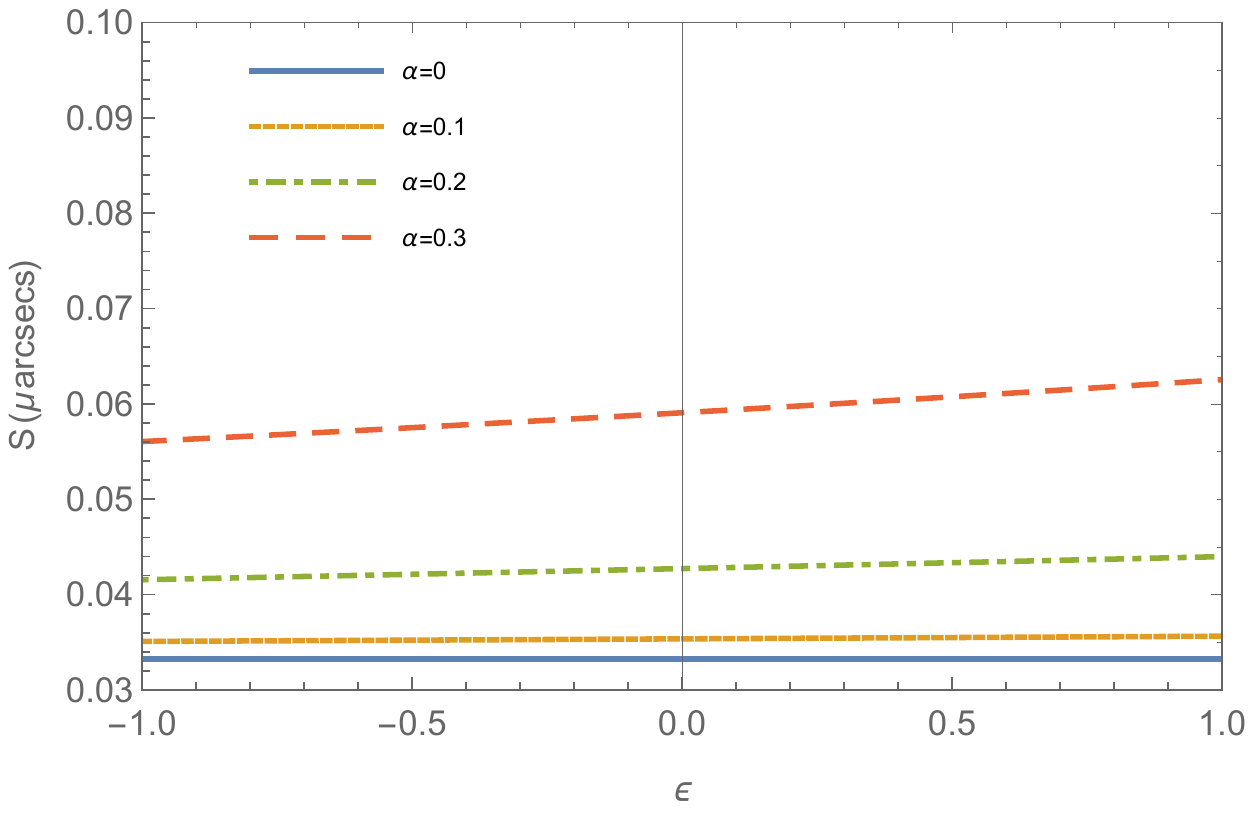}}\\
	\subfloat[$r_m=2.5\log_{10}\tilde{r}$]{\includegraphics[width = 3in]{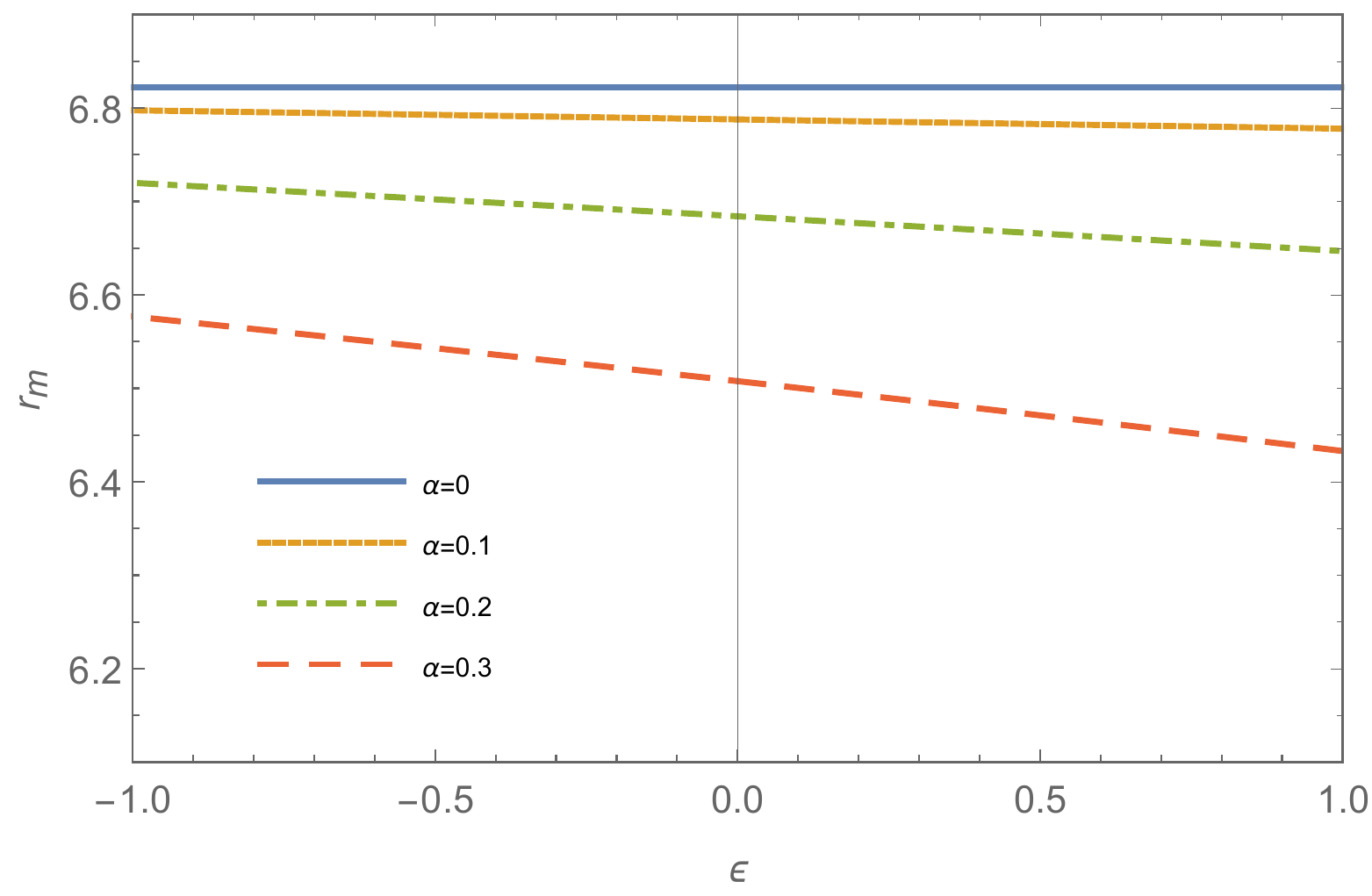}} 
	\caption{observable}
	\label{obser}
\end{figure}

In Fig. \ref{LG}, the light source, located in {\bf S}, emits radiation deflected towards the observer {\bf O} due to the presence of the compact object {\bf L}. We represent the angular position of the source by $\beta$, the angular position of the image measured by the observer by $\theta$, and the angular deflection of light by $\Lambda$. In addition, $D_{LS}$ is the distance between the source {\bf S} and the lens {\bf L},  $D_{OL}$ is the distance between the observer {\bf O} and the lens {\bf L} e $D_{OS}=D_{OL}+D_{LS}$ is the distance between the observer and the source. All these quantities are measured with respect to the optical axis, which is the straight line through {\bf OL}. As in \cite{Boz-Cap2001}, we assume that the source {\bf S}  is almost perfectly aligned with the lens {\bf L}; in this limit, the relationship between the angular position of the source, the position of the image seen by the observer  and the angular deflection  is given by
\begin{equation}\label{Eq.L}
\beta=\theta-\frac{D_{LS}}{D_{OS}}\Delta\Lambda_{n}\ ,
\end{equation}
where  $\Delta\Lambda_{n}$ is the deflection angle with subtracting all the loops performed by the photons before moving towards the observer, that is, $\Delta\Lambda_{n}=\Lambda-2n\pi$. Under this condition of alignment, one finds $b\simeq\theta D_{OL}$. Thus, the deflection angle is given by

\begin{equation}\label{defle}
\Lambda(\theta)=-\bar{a}\log\left(\frac{\theta D_{OL}}{b_c}-1\right)+\bar{b}\ .
\end{equation}
To find $\Delta\Lambda_{n}$  entering the lens equation, we expand $\Lambda (\theta)$ around $\theta=\theta_n^0$ (where $\Lambda(\theta^0_n)=2n\pi$): $\Delta\Lambda_n=\frac{\partial\Lambda}{\partial\theta}|_{\theta=\theta^0_n}(\theta-\theta^0_n)$. From evaluating (\ref{defle}) in $\theta=\theta_0$, it follows that
\begin{equation}
\theta^0_{n}=\frac{b_c}{D_{OL}}\left(1+e_n\right), \qquad e_n=e^{\bar{b}-2n\pi} \ .
\end{equation}
Thus, we obtain that $\Delta\Lambda_n=-\frac{\bar{a}D_{OL}}{b_ce_n}(\theta-\theta^0_n)$. Introducing these results in the lens equation (\ref{Eq.L}) and noting that $\frac{b_c}{D_{OL}}\ll1$, we obtain the position of the $n^{\text{th}}$  relativistic image
\begin{equation}
\theta_n\simeq\theta^0_n+\frac{b_ce_n}{\bar{a}}\frac{D_{OS}}{D_{OL}D_{LS}}(\beta-\theta^0_n) \ .
\end{equation}
Another observable of interest is the magnification $\mu_{n}$ which is given by  inverse of
the modulus of the Jacobian, $	\mu_n=\left|\frac{\beta}{\theta}\frac{\partial\beta}{\partial\theta}|_{\theta^0_{n}}\right|^{-1}$:
\begin{equation}
\mu_{n}=\frac{e_n(1+e_n)}{\bar{a}\beta}\frac{D_{OS}}{D_{LS}}\left(\frac{b_c}{D_{OL}}\right)^2 \ .
\end{equation}
It is worth stressing out that we express the positions of the relativistic images and also the magnification in terms of the coefficients of the expansion. Once having found them one can compare with the experimental data.
The inverse problem consist of finding the coefficients of the expansion in the strong field limit from the positions and the flow, and thus to  discover the nature of the object responsible for the lens. For this, we suppose that only the outermost image $\theta_1$ is discriminated, while all others are encompassed in $\theta_\infty$.
Our observables are \cite{Bozza2002}
\begin{eqnarray}
\theta_{\infty}&=&\frac{b_c}{D_{OL}},\\
s&=&\theta_{1}-\theta_{\infty}= \theta_{\infty} e^{\frac{\bar{b}-2\pi}{\bar{a}}},\\
\tilde{r}&=&e^{\frac{2\pi}{\bar{a}}} \ .
\end{eqnarray}
Here $s$ is angular separation and $\tilde{r}=\frac{\mu_1}{\sum\limits_{n=2}^{\infty} \mu_{n}}$ is the relationship between the flux of the first image and the flux of all other images. To perform the analysis of the observables, let us consider that the gravitational lens is derived from a black hole, like the one in the center of our galaxy, Milky Way, and that the geometry of the spacetime is described by the metric (\ref{mtr}). The mass of the black hole is estimated as $4.4 \times10^6 M_{\odot}$ and its distance to the earth is approximately $D_{OL}=8,5$kpc \cite{Genzel}. With this, we estimate the values of the observables  and plot the Fig.\ref{obser}. We observe that the position $\theta_{\infty}$ of the relativistic image and the relative magnification $r_{m}$ ($r_m=2.5\log_{10}\tilde{r}$) decrease as $\epsilon$ increases. The angular separation $s$  increases as $\epsilon$ increases. Regarding the topological charge, both $\theta_{\infty}$ and  $s$ grow as $\alpha$ increases, this behavior also occurs in GR \cite{Man2011}. From the Fig. \ref{obser} and the table \ref{t}, we can extract  the following information: for a given value of $\alpha$, we observe that $\theta_{\infty}^{GM}>\theta_{\infty}^{EiBI}>\theta_{\infty}^{Sch} $ to $\epsilon>0$, and $\theta_{\infty}^{EiBI}>\theta_{\infty}^{GM}>\theta_{\infty}^{Sch} $ to $\epsilon<0$. As for $s$, we have $s^{EiBI}>s^{GM}>s^{Sch}$ to $\epsilon>0$, and $s^{GM}>s^{EiBI}>s^{Sch}$ to $\epsilon<0$. As for $r_m$, for a given  $\alpha$, we have $r_m^{Sch}>r_m^{GM}>r_m^{EiBI}$ to $\epsilon>0$, and $r_m^{Sch}>r_m^{EiBI}>r_m^{GM}$ to $\epsilon<0$. Note that the observables related to a black hole with topological charge in the GR and in the EiBI gravity  exchange mutually when we go from $\epsilon>0$ to $\epsilon<0$. 

\begin{table}[!ht]\small
	\caption{Observable}
	\label{t}
	\begin{tabular}{|c |c |c |c| c| }
		\hline
		\textbf{$\epsilon$}                                                              &\textbf{ $\alpha$} & \textbf{$\theta_{\infty}$($\mu$ arcsecs)} & \textbf{s($\mu$ arcsecs)} & \textbf{$r_m$(magnitudes)} \\ \hline
		Schw                                                                    & 0        & 26.5473           & 0.03322          & 6.8218            \\ \hline
		\multirow{3}{*}{GM}                                                     & 0.1      & 26.9506           & 0.03535          & 6.7876            \\ \cline{2-5} 
		& 0.2      & 28.2237           & 0.04272          & 6.6840            \\ \cline{2-5} 
		& 0.3      & 30.5815           & 0.05907          & 6.5076            \\ \hline
		\multirow{3}{*}{\begin{tabular}[c]{@{}c@{}}EiBI\\ \\ -0.3\end{tabular}} & 0.1      & 26.9682           & 0.03527          & 6.79064           \\ \cline{2-5} 
		& 0.2      & 28.2929           & 0.04237          & 6.69494           \\ \cline{2-5} 
		& 0.3      & 30.7326           & 0.05812          & 6.52897           \\ \hline
		\multirow{3}{*}{\begin{tabular}[c]{@{}c@{}}EiBI\\ \\ -0.2\end{tabular}} & 0.1      & 26.9623           & 0.03530          & 6.78966           \\ \cline{2-5} 
		& 0.2      & 28.2698           & 0.04248          & 6.69132           \\ \cline{2-5} 
		& 0.3      & 30.6824           & 0.05843          & 6.52192           \\ \hline
		\multirow{3}{*}{\begin{tabular}[c]{@{}c@{}}EiBI\\ \\ -0.1\end{tabular}} & 0.1      & 26.9564           & 0.03533          & 6.78867           \\ \cline{2-5} 
		& 0.2      & 28.2468           & 0.04260          & 6.6877            \\ \cline{2-5} 
		& 0.3      & 30.632            & 0.05875          & 6.51482           \\ \hline
		\multirow{3}{*}{\begin{tabular}[c]{@{}c@{}}EiBI\\ \\ 0.1\end{tabular}}  & 0.1      & 26.9447           & 0.03538          & 6.7867            \\ \cline{2-5} 
		& 0.2      & 28.2005           & 0.04285          & 6.68039           \\ \cline{2-5} 
		& 0.3      & 30.5307           & 0.05940          & 6.50045           \\ \hline
		\multirow{3}{*}{\begin{tabular}[c]{@{}c@{}}EiBI\\ \\ 0.2\end{tabular}}  & 0.1      & 26.9388           & 0.03541          & 6.78571           \\ \cline{2-5} 
		& 0.2      & 28.1774           & 0.04297          & 6.67672           \\ \cline{2-5} 
		& 0.3      & 30.4798           & 0.05973          & 6.49318           \\ \hline
		\multirow{3}{*}{\begin{tabular}[c]{@{}c@{}}EiBI\\ \\ 0.3\end{tabular}}  & 0.1      & 26.9329           & 0.03544          & 6.78473           \\ \cline{2-5} 
		& 0.2      & 28.1541           & 0.04309          & 6.67304           \\ \cline{2-5} 
		& 0.3      & 30.4286           & 0.06006          & 6.48585           \\ \hline
	\end{tabular}
\end{table}

 \section{DEFLECTION ANGLE IN THE ELLIS WORMHOLE SPACETIME WITH TOPOLOGICAL CHARGE}\label{5}
When the integration constant $M_0$ is chosen to be equal to zero, in the case $\epsilon<0$ in (\ref{wormhole}), the solution describes a traversable wormhole with a topological charge $\alpha$,
 \begin{equation}\label{mm}
 	ds^2=-dt^2+dx^2+(1-\alpha^2)(x^2+r_0^2)(d\theta^2+\sin^2\theta d\phi^2);
 \end{equation}
 where $r_0=\sqrt{|\epsilon|}\alpha$ is the radius of the throat of the wormhole. As in \cite{Tsukamoto2017}, let us focus on the side $x>0$ and admit that the light ray does not go through the throat of the wormhole. From (\ref{rps}), we conclude that the radius of the photon sphere is $x_m=0$. But $x_m$  is assumed to be positive. Thus we introduce a new radial coordinate $r$, given by 
 \begin{equation}
 	r=x+p, \qquad\text{where $p>0$ is a constant}.
 \end{equation}
 Thus, the (\ref{mm}) becomes
 \begin{equation}
 	ds^2=-dt^2+dr^2+(1-\alpha^2)\left[(r-p)^2+r_0^2\right](d\theta^2+\sin^2\theta d\phi^2).
 \end{equation}
Therefore,
\begin{eqnarray}
	A&=&1,\\
	B&=&1,\\
	C&=&(1-\alpha^2)\left[(r-p)^2+r_0^2\right].
\end{eqnarray} 
In terms of this new radial coordinate, the radius of the photon sphere, the critical parameter $b_c$,  $\bar{a}$ and $\bar{b}$ are given by
 \begin{eqnarray}
 	r_m&=&p,\\
 	b_c&=&r_0\sqrt{1-\alpha^2},\\
 	\bar{a}&=&\frac{1}{\sqrt{1-\alpha^2}},\\
 	\bar{b}&=& \frac{1}{\sqrt{1-\alpha^2}}\log\left(\frac{2p^2}{r_0^2}\right)+I_R(r_m)-\pi,
 \end{eqnarray} 
 From  (\ref{regp}), we have
 \begin{eqnarray}
 	I_R(r_m)&=&\frac{2}{\sqrt{1-\alpha^2}}\int_{0}^{1}\left(\frac{r_0}{z\sqrt{r_0^2-2r_0^2z+(r_0^2+p^2)z^2}}-\frac{1}{z}\right)dz,\nonumber\\
 	I_R(r_m)&=&\frac{2}{\sqrt{1-\alpha^2}}\log\left(\frac{2r_0}{p}\right).
 \end{eqnarray}
 Using all this, we find $\bar{b}=\frac{3}{\sqrt{1-\alpha^2}}\log(2)-\pi$. Thus, the deflection angle (\ref{angdefle}) takes the form
 \begin{equation}\label{Whd}
 	\Lambda(b)=-\frac{1}{\sqrt{1-\alpha^2}}\log\left(\frac{b}{r_0\sqrt{1-\alpha^2}}-1\right)+\frac{3}{\sqrt{1-\alpha^2}}\log 2-\pi +\mathcal{O}[(b-b_c)\log(b-b_c)].
 \end{equation}
As we can see, the presence of the topological charge increases the expansion coefficients for the angular deflection at the strong field limit. As expected, taking $\alpha=0$ and assuming  $r_0\neq|\epsilon|\alpha$, this deflection reduces to that of Ellis wormhole, in the GR case \cite{Tsukamoto2016} \footnote{We also observe that this result is in agreement with  \cite{HSR}, the authors evaluated the light deflection in a topologically charged wormhole background 
 (\ref{mm}). They have used the strong field limit  $\lim\limits_{g\to1}K(g)=-\frac{1}{2}\log(1-g)+\frac{3}{2}\log2+\mathcal{O}[(1-g)\log(1-g)]$, check  Eq.(9) from  \cite{HSR} for more details.}.
 
 \section{SUMMARY AND CONCLUSIONS}\label{6}
 In this paper, we considered the gravitational lensing in EiBI gravity in a space-time with a topological charge.
Starting from the energy-momentum tensor corresponding to the region external to the GM core, we reproduced the solution first obtained in \cite{lambaga2018}. We followed the approach developed in \cite{olmo2011} and found that, depending on the value of the EiBI parameter $\epsilon$, the solution can describe both a black hole and a traversable wormhole with the topological charge  of the GM ($\epsilon<0$).  In relation to GR, $\epsilon>0$  has the effect of decreasing the event horizon radius of the corresponding black hole, while $\epsilon<0$ increases. When we take $\epsilon=0$ or $\alpha=0$, which corresponds to the vacuum solution, we return to the GR, as expected. In addition, asymptotically the solution we found tends to GR situation. Therefore, it was necessary to investigate the deflection of light in the strong field limit,  as a counterpart to \cite{lambaga2018}, where the deflection of light was studied, but in the asymptotic regime.
 
We adopted the methodology developed by Bozza \cite{Bozza2002} and improved by Tsukamoto \cite{Tsukamoto2017} to get the expansion  of the angle of deflection of light in the strong field limit  of our solution. Initially we considered the solution of black holes taking into account both cases, $\epsilon>0$ and
$\epsilon<0$, in the region $x>0$. We obtained the coefficients of the expansion and expressed them analytically up to the first order in the EiBI parameter $\epsilon$, maintaining the exact expression for the GR part.  We have seen that, for a fixed value of $\alpha$, with the exception of the critical impact parameter, all expansion coefficients, including the angular deflection  $\Lambda(\theta)$, given by Fig. \ref{CE}, increase as $\epsilon$ grows. As for $\alpha$, all expansion coefficients increase while $\alpha$ grows \cite{Man2011}. At the next step, we numerically evaluated the observables $\theta_{\infty}$, $s$ and $r_m$, allowing to measure important characteristics of relativistic images. For this, we simulated a scenario in which the object would be the black hole in the center of our galaxy, and the geometry is given by (\ref{metrica-x}). We then find that the position of the relativistic images $\theta_{\infty}$ and the relative magnification $r_m$ decrease with $\epsilon$ in comparison to the GR, while $s$ increases. 

We still consider the strong field lensing due the presence of the to wormhole with a topological charge. Using the approach developed in \cite{Tsukamoto2017}, we have analytically obtained the expansion for the deflection angle, showing its dependence on $\alpha$ and $\epsilon$ (\ref{Whd}).

In short, we have seen that the strong dependence of the observables on the parameters of the object, either a black hole or a wormhole, allows to distinguish the results predicted by GR and EiBI gravity in the strong field limit. Therefore, we can test the possibility of deviations from GR in the regime of high energies arising due to the topological charge of the GM.

 \appendix
 \section{RADIUS OF THE PHOTON SPHERE}\label{ap}
  The radius of the photon sphere is given by the greatest of the real roots of the equation
   \begin{equation}\label{ce1}
   x^3-\frac{3x^2}{2(1-\alpha^2)}+\frac{\epsilon\alpha^2}{2(1-\alpha^2)}=0.
   \end{equation}
  Following \cite{Hbook}, for simplicity, we can write this equation as
  \begin{equation}
  	x^3+a_1x^2+a_2=0,
  \end{equation}
with the coefficients $a_1=-\frac{3}{2(1-\alpha^2)}$ e $a_2=\frac{\epsilon\alpha^2}{2(1-\alpha^2)}$. In terms of the following quantities
\begin{equation}
	Q=-\frac{a_1^2}{9}\qquad \text{and}\qquad R=\frac{-2a_1^3-27a_2}{54},
\end{equation}  
 the discriminant $D$ of the equation is given by  $D=Q^3+R^2$, that is,
 \begin{equation}
 	D= \left(\frac{\epsilon\alpha^2}{4(1-\alpha^2)^2}\right)^2\left[(1-\alpha^2)^2-\frac{1}{\epsilon\alpha^2}\right].
 \end{equation}
 When $\epsilon>0$, the condition (\ref{rest}) for avoiding naked singularities yields $\frac{1}{\epsilon\alpha^2}\ge(1-\alpha^2)^2$. Since we have no interest in the transient case, we have $D<0$, and thus (\ref{ce1}) has three real solutions, the largest of which is given by
 \begin{equation}
 	x_m=2\sqrt{-Q}\cos\left[\frac{1}{3}\arccos\left(\frac{R}{\sqrt{-Q^3}}\right)\right]-\frac{a_1}{3}.
 \end{equation}
 Substituting their values, we have
 \begin{equation}\label{e1}
 	x_m=\frac{1}{2(1-\alpha^2)}+\frac{1}{(1-\alpha^2)}\cos\left[\frac{1}{3}\arccos\left(1-2\epsilon\alpha^2(1-\alpha^2)^2\right)\right]\quad(\epsilon>0).
 \end{equation}
 We conclude that the radius of the photon sphere decreases when $\epsilon>0$ increases.
 
 In the case $\epsilon<0$, the discriminant is positive, $D>0$, then there is only one real root for the equation (\ref{ce1}), which is given by
 \begin{equation}
 	x_m=\left[R+\sqrt{D}\right]^{1/3}+\left[R-\sqrt{D}\right]^{1/3}-\frac{a_1}{3} \ ,
 \end{equation}
 that is,
 
 \begin{eqnarray}\label{e2}
 x_m&=&\frac{1}{2(1-\alpha^2)}+\left[\frac{1}{8(1-\alpha^2)^3}-\frac{\epsilon\alpha^2}{4(1-\alpha^2)}+\frac{\epsilon\alpha^2}{4(1-\alpha^2)^2}\sqrt{(1-\alpha^2)^2-\frac{1}{\epsilon\alpha^2}}\right]^{1/3}+\nonumber\\
 &+&\left[\frac{1}{8(1-\alpha^2)^3}-\frac{\epsilon\alpha^2}{4(1-\alpha^2)}-\frac{\epsilon\alpha^2}{4(1-\alpha^2)^2}\sqrt{(1-\alpha^2)^2-\frac{1}{\epsilon\alpha^2}}\right]^{1/3} \quad(\epsilon<0).
 \end{eqnarray}
We can expand $x_m$ in series of powers in both cases (\ref{e1}) and (\ref{e2}) to obtain the leading order
\begin{equation}\label{xapp}
	x_{m}\simeq\frac{3}{2(1-\alpha^2)}-\frac{2\epsilon\alpha^2}{9},
\end{equation}
this applies to both cases $\epsilon>0$ and  $\epsilon<0$.

\textbf{Acknowledgments.}  
Authors are grateful to Gonzalo Olmo for helpful discussions and collaboration on related topics. The work by A. Yu. P. has been supported by the
CNPq project No. 301562/2019-9. P. J. Porf\'{i}rio would like to acknowledge the Brazilian agency CAPES for the financial support, and also the Departament de F\'{i}sica Te\`{o}rica and IFIC, Universitat de Val\`{e}ncia, for the hospitality.

\end{document}